\documentclass[twocolumn,           
               showpacs,            
               preprintnumbers,     
               aps,                 
               prd,          	    
               a4paper,             
               superscriptaddress,  
               unsortedaddress,     
               footinbib,           
               tightenlines,        
               floats               
               ]{revtex4}

\usepackage{graphicx}
\usepackage{latexsym}
\usepackage{amsmath,amssymb}        
\usepackage[draft=false]{hyperref}

\newcommand{\de}{\mathrm{d}}
\renewcommand{\(}{\left(}
\renewcommand{\)}{\right)}
\renewcommand{\[}{\left[}
\renewcommand{\]}{\right]}
\newcommand{\period}{\,\mathrm{.}}
\newcommand{\comma}{\,\mathrm{,}}
\newcommand{\reffig}[1]{Fig.~\ref{#1}}
\newcommand{\refeq}[1]{Eq.~(\ref{#1})}

\newcommand{\mpl}{m_\mathrm{Pl}}

\newcommand{\abs}[1]{\left\vert#1\right\vert}

\newcommand{\pti}{\tilde{p}}
\newcommand{\eti}{\tilde{\varepsilon}}
\newcommand{\wk}{w_\mathrm{k}}
\newcommand{\wf}{w_\mathrm{f}}
\newcommand{\csk}{c^2_\mathrm{sk}}
\newcommand{\Qti}{\tilde{Q}}
\newcommand{\Vti}{\tilde{V}}
\newcommand{\Uti}{\tilde{U}}

\begin{document}


\title{A new view of k-essence}
\author{Micha\"el Malquarti}
\affiliation{Astronomy Centre, University of Sussex, 
             Brighton BN1 9QJ, United Kingdom}
\author{Edmund J. Copeland}             
\affiliation{Centre for Theoretical Physics, University of Sussex, 
             Brighton BN1 9QJ, United Kingdom}
\author{Andrew R. Liddle}
\affiliation{Astronomy Centre, University of Sussex, 
             Brighton BN1 9QJ, United Kingdom}
\author{Mark Trodden}
\affiliation{Department of Physics, Syracuse University,
              Syracuse, NY 13244-1130, USA}
\date{\today} 
\pacs{98.80.Cq \hfill astro-ph/0302279}
\preprint{astro-ph/0302279}


\begin{abstract}
K-essence models, relying on scalar fields with non-canonical kinetic terms, 
have been proposed as an alternative to quintessence in explaining the observed 
acceleration of the Universe. We consider the use of 
field redefinitions to cast k-essence in a more familiar form. While k-essence 
models cannot in general be rewritten in the form of quintessence models, we 
show that in certain dynamical regimes an equivalence can be made, which in 
particular can shed light on the tracking behaviour of k-essence. In several 
cases, k-essence cannot be observationally distinguished from quintessence using 
the homogeneous evolution, though there may be small effects on the perturbation 
spectrum. We make a detailed 
analysis of two k-essence models from the literature and comment on the nature 
of the fine tuning arising in the models.
\end{abstract}

\maketitle


\section{Introduction}

One of the greatest challenges in modern cosmology is understanding the nature 
of the dark energy responsible for the observed acceleration of the present 
Universe~\cite{acce}. A popular framework, known as quintessence, involves 
scalar 
field models in which the field slow-rolls down a potential, with its potential 
energy acting analogously to that of early Universe inflation~\cite{quin,trac} 
models. However, recently a second possibility, that of an effective scalar 
field theory described by a Lagrangian with a non-canonical kinetic term, has 
also been proposed. Such a model may lead to early time acceleration, where it 
is named k-inflation~\cite{ADM,GM}, or acceleration in the present Universe 
under the name k-essence~\cite{COY,AMS1,AMS2}. It is worth noting that tachyon 
dark energy models~\cite{tach} may be seen as special cases of k-essence.

Allowing the dark energy to be dynamical provides an opportunity to study the 
so-called coincidence problem, which asks why dark energy domination begins just 
at the epoch when we cosmologists exist to observe it. Traditional quintessence 
models appear promising in this regard, as they can support scaling or tracking 
solutions, in which the scalar field energy density follows that of the dominant 
source of matter~\cite{trac}. Unfortunately, these models require a fine-tuning 
of parameters, making them not particularly more attractive than a pure 
cosmological constant in the cases considered so far. Alternatively, a class of 
k-essence models~\cite{AMS1,AMS2} has been claimed to solve the coincidence 
problem in a generic way; after a long period of perfect tracking, the 
domination of dark energy is triggered by the transition to matter domination, a 
time period during which structures --- and cosmologists --- can form. 
Therefore, it is important to understand the extent to which these models differ 
from quintessence models. In particular, one would like to understand whether 
these models suffer from the same sort of fine tuning issues as quintessence, 
and also whether observations can differentiate between the two models. We will 
address this question by employing field redefinitions which allow k-essence 
models to be recast in a form similar to quintessence, though we stress 
immediately that in general the two ideas are distinct and it is not possible to 
write an arbitrary k-essence model in quintessence form.

Throughout this article a prime denotes a derivative with respect to the 
argument of the function to which it is applied, and a dot denotes a derivative 
with respect to time.

\section{K-essence}
\label{k_essence}

To begin with, we wish to be absolutely clear in our terminology. Although the 
literature contains different usages, particularly of the word quintessence, it 
is important that our own usage be unambiguous. We will use the word 
`quintessence' exclusively to refer to models which feature a single scalar 
field with a canonical kinetic term (whereas in some other papers the definition 
of 
quintessence is so general as to include k-essence within it). 

We now introduce the k-essence model. Neglecting for now the part of the 
Lagrangian containing ordinary matter, the action for a k-essence field $\phi$ 
is given by
\begin{equation}
\label{k_action}
S=\int\de^4x\sqrt{-g}\[-\frac{\mpl^2}{16\pi}R+K(\phi)\pti(X)\]\comma
\end{equation}
where we assume $K(\phi)>0$ and $X=\frac12\nabla_\mu\phi\nabla^\mu\phi$. The 
name of the model suggests that the field should be driven only by its kinetic 
energy. For that to be strictly true one should impose $\pti(0)=0$, otherwise 
that term could be separated as a potential term independent of 
$\nabla_\mu\phi$. Indeed, for sufficiently small $X$, one could even write 
$\pti(X)\approx\pti(0)+\pti'(0)X$ and, after field redefinition, obtain a 
canonical scalar field with a potential. However this condition is not imposed 
in most k-essence papers, e.g.~Refs.~\cite{AMS1,AMS2}, and we will not impose it 
either.

In order to be even more general, one may include a separate potential term 
$\bar{U}(\phi)$ in the Lagrangian, now setting $\pti(0)=0$~\cite{MMOT}. 
Obviously, this model includes the one given by \refeq{k_action}. We give an 
extension of our results to this more general possibility in the appendix.

Now, we recall some properties summarized in Ref.~\cite{AMS2}. Using the perfect 
fluid analogy, the pressure and the energy density are given by
\begin{eqnarray}
p &=& K(\phi)\pti(X)\comma\\
\varepsilon &=& K(\phi)\eti(X)\comma
\end{eqnarray}
where
\begin{equation}
\label{eps_ti}
\eti(X)=2X\pti'(X)-\pti(X)\period
\end{equation}
The equation of state parameter is given by
\begin{equation}
\wk=\frac{\pti(X)}{\eti(X)}=\frac{\pti(X)}{2X\pti'(X)-\pti(X)}\comma
\end{equation}
while the effective sound speed is given by
\begin{equation}
\label{sound_speed}
\csk=\frac{\pti'(X)}{\eti'(X)}
    =\frac{\pti'(X)}{\pti'(X)+2X\pti''(X)}\period
\end{equation}
This definition comes from the equation describing the evolution of linear 
perturbations in a k-essence dominated Universe~\cite{GM}, and therefore is 
relevant when studying the stability of the theory. We note, however, that, as 
shown in Ref.~\cite{CHT}, $\csk>0$ is not a sufficient condition for the theory 
to be stable. It is important to notice that the effective sound speed is not 
always equal to $1$, as it is in quintessence models. Therefore, the behaviour 
of perturbations in this case is genuinely different from that in the case of 
canonical scalar fields and this difference may be observable~\cite{sosp}. Also, 
note that the definition given in \refeq{sound_speed} is different from the 
thermodynamic definition of the isentropic sound speed
\begin{equation}
c^2_\mathrm{s}=\(\frac{\partial p}{\partial\varepsilon}\)_S
              =\frac{\dot{p}}{\dot{\varepsilon}}\period
\end{equation}
As explained in Ref.~\cite{KS}, the difference arises from the fact that ``a 
well-defined concept of sound speed does not exist for classical scalar fields'' 
and therefore their density perturbations behave quite differently from the 
usual hydrodynamic case.

Although models with negative energy density, super-negative or diverging 
equation of state and/or imaginary or diverging sound speed have been studied, 
e.g.~Refs.~\cite{COY,AMS1,AMS2}, it is possible to restrict the class of models 
by imposing one or more of the following independent constraints:
\begin{equation}
\begin{array}{lcl}
\varepsilon>0 &\Rightarrow& 2X\pti'(X)>\pti(X)\comma\\
\wk>-1 &\Rightarrow& \eti(X)\pti'(X)>0\comma\\
\csk>0 &\Rightarrow& 2X\pti''(X)\pti'(X)>-\pti'^2(X)\period
\end{array}
\end{equation}

{}From the action Eq.~(\ref{k_action}), in the case of a flat Robertson--Walker 
metric
\begin{equation}
\label{frwmetric}
\de s^2=-\de t^2+a^2(t)\de\mathbf{x}^2\comma
\end{equation}
the Euler--Lagrange equation for the k-essence field is
\begin{equation}
\eti'(X)\ddot{\phi}+3H\pti'(X)\dot{\phi}+\frac{K'(\phi)}{K(\phi)}\eti(X)=0\comma
\end{equation}
where $H=\dot{a}/a$. We see that if $\eti'(X)=0$ at some $X_\mathrm{c}$, so that 
$\csk$ diverges, the equation is singular and reduces to a first-order equation 
which gives a constraint on $\phi$. Some problems may arise at this singularity, 
but we leave this issue for future investigation. In any case, regions 
separated by a diverging sound speed are disconnected and may be considered as 
different models. In each such region the Euler--Lagrange equation may be 
rewritten as
\begin{equation}
\ddot{\phi}+3H\csk(X)\dot{\phi}+\frac{K'(\phi)}{K(\phi)}\frac{\eti(X)}{\eti'(X)}
=0\period
\end{equation}

Finally, note that, for models in which $\pti(X)$ is analytic and equal to $0$ 
at the origin, then if $\pti'(X=0)<0$ the function $\eti(X)$ must be negative 
for some $X$. Given that $\pti(X)$ is analytic at the origin, we then have
\begin{equation}
\pti(X)=-p_nX^n+\mathcal{O}(X^{n+1})\comma
\end{equation}
which implies
\begin{equation}
\eti(X)=-p_n(2n-1)X^n+\mathcal{O}(X^{n+1})\comma
\end{equation}
with $n$ a positive integer and where $p_n>0$ is the first non-zero 
co-efficient. Hence $\eti<0$ for some range of $X$.

In the models described in Ref.~\cite{AMS2}, $\pti(X)$ is negative for small $X$ 
and therefore, if $\pti(0)=0$ then $\pti'(X)$ must be negative close to the 
origin. Hence $\eti(X)<0$ over some range of $X$. As a result, for these models, 
if one imposes $\varepsilon(X)>0$ and the analyticity of $\pti(X)$ at the 
origin, then $\pti(0)$ must be non-zero and, as explained above, this implies 
that one has, in effect, introduced a potential term.

\section{Dynamical equivalence}
\label{dyn_equiv}

In this section we present a model which is dynamically equivalent to the 
k-essence model. Consider a new action
\begin{equation}
\label{newaction}
\begin{split}
S=&\int\de^4x\sqrt{-g}\bigg\{-\frac{\mpl^2}{16\pi}R\\
&+K(\phi)\Big[\pti(\chi)+\big(\frac12\nabla_\mu\phi\nabla^\mu\phi
-\chi\big)\pti'(\chi)\Big]\bigg\}\comma
\end{split}
\end{equation}
in which we have introduced a field $\chi$, which acts like a Lagrange 
multiplier. The variational principle with respect to $\chi$ gives
\begin{equation}
\pti''(\chi)\(\chi-\frac12\nabla_\mu\phi\nabla^\mu\phi\)=0\comma
\end{equation}
and therefore --- as long as $\pti''(\chi)\neq0$ --- the 
action Eq.~(\ref{newaction}) is dynamically equivalent to Eq.~(\ref{k_action}). 
Note, 
however, that, when quantized, the theories will no longer be equivalent. Now 
perform a field transformation
\begin{equation}
Q\equiv\int_{\phi_0}^\phi\sqrt{K(\sigma)} \, \de\sigma\comma
\end{equation}
which leads to
\begin{equation}
\begin{split}
S=&\int\de^4x\sqrt{-g}\bigg\{-\frac{\mpl^2}{16\pi}R\\
  &+\frac{\pti'(\chi)}{2}\nabla_\mu Q\nabla^\mu Q
   +V(Q)\big[\pti(\chi)-\chi\pti'(\chi)\big]\bigg\}\comma
\end{split}
\end{equation}
where $V(Q)\equiv K[\phi(Q)]$. From now on, for simplicity, we assume that 
$\pti''(X)$ has a constant sign, although this is not a particularly strong 
assumption. This allows us to define a new field by computing a modified 
Legendre transformation
\begin{eqnarray}
\psi &\equiv& \pti'(\chi)\comma\\
W(\psi) &\equiv& \chi\pti'(\chi)-\pti(\chi)\comma
\end{eqnarray}
which finally gives
\begin{equation}
\begin{split}
S=&\int\de^4x\sqrt{-g}\bigg[-\frac{\mpl^2}{16\pi}R\\
  &+\frac\psi2\nabla_\mu Q\nabla^\mu Q-V(Q)W(\psi)\bigg]\period
\end{split}
\end{equation}

This action is very simple, but contains a non-dynamical field $\psi$ coupled 
non-canonically to a canonical scalar field $Q$. Clearly, if we impose 
$\varepsilon>0$, then the condition $\wk>-1$ implies that $\psi$ is positive, 
and therefore that the kinetic term has the canonical sign. In the opposite 
case, the kinetic term has the sign of the phantom model introduced by 
Caldwell~\cite{phan}.

For the flat Robertson--Walker metric~(\ref{frwmetric}), the energy density and 
pressure of this two-field component of matter are
\begin{eqnarray}
\rho_\mathrm{Q\psi} &=& \frac\psi2\dot{Q}^2+V(Q)W(\psi)\comma\\
p_\mathrm{Q\psi} &=& \frac\psi2\dot{Q}^2-V(Q)W(\psi)\period
\end{eqnarray}
The Euler--Lagrange equation for $Q$ is given by
\begin{equation}
\psi\ddot{Q}+(3H\psi+\dot{\psi})\dot{Q}+V'(Q)W(\psi)=0\comma
\end{equation}
and the constraint equation for $\psi$ is given by
\begin{equation}
\label{constraint}
W'(\psi)=\frac{\dot{Q}^2}{2V(Q)}\period
\end{equation}

If $\psi$ is positive and remains almost constant, the field $Q$ plays the role 
of a canonical scalar field. Indeed, by renormalizing $Q$ as $\Qti=\sqrt{\psi}Q$ 
we obtain an (almost) equivalent quintessence model which should mimic the 
k-essence field during the time period for which the assumption of approximately 
constant $\psi$ holds. If this is the case when k-essence starts dominating, it 
will not be possible to distinguish it observationally from a quintessence 
field, unless one takes into account perturbations. As described in 
Refs.~\cite{COY,AMS1,AMS2} the k-essence field can undergo several attractor 
regimes during which its kinetic energy remains constant. Hence, $\psi$ is 
indeed constant over those periods of time. Thus, during some time periods, the 
homogeneous part of the k-essence field will behave exactly like a quintessence 
field.

\section{Examples}
\label{examples}

In this section we study two examples of k-essence models. We first study the 
model described in Ref.~\cite{COY} and given by
\begin{eqnarray}
K(\phi)&=&\(\frac{\phi}{\mpl}\)^{-\alpha}\comma\\
\pti(X)&=&P_0\[-\frac{X}{\mpl^4}+\(\frac{X}{\mpl^4}\)^2\]\period
\end{eqnarray}
It is easy to check that this is a rather unconventional model; for some ranges 
of $X$ we can have $\varepsilon<0$, $\wk<-1$ or $\csk<0$, and $\wk$ and $\csk$ 
diverging at $X_w=\mpl^4/3$ and $X_c=\mpl^4/6$ respectively. Nevertheless, 
this model features an interesting behaviour --- in the presence of a dominating 
fluid with equation of state parameter $\wf$ there exists a stable scaling 
solution for which $X$ is constant and
\begin{equation}
\wk=\frac{(1+\wf)\alpha}{2}-1\period
\end{equation}

\begin{figure}[t]
\includegraphics[scale=0.38,angle=-90]{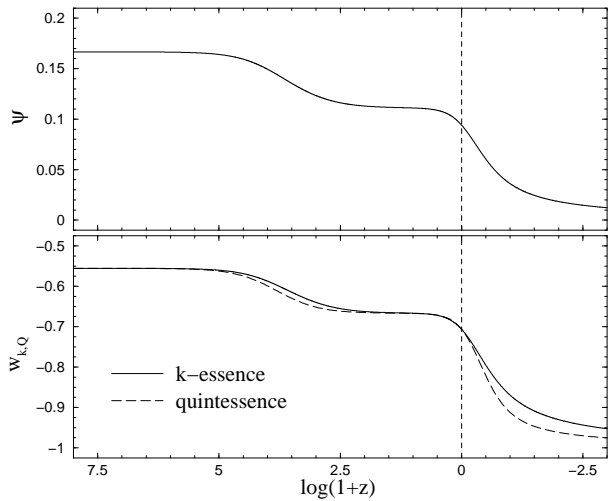}
\caption{Top panel: Evolution of $\psi$ for the first example of a k-essence 
field described in Section~\ref{examples} in a cosmologically realistic case and 
with $\alpha=2/3$. Note the transitions from radiation domination to matter 
domination and from matter domination to k-essence domination. Bottom panel: 
Evolution of the equation of state parameter $w$ for this field and for the 
almost equivalent quintessence field, that is to say an inverse power-law model 
with $\beta=1$.}
\label{tracker}
\end{figure}

Using the transformation described in the last section, we find
\begin{equation}
V(Q)=(\beta/\alpha)^\beta\(\frac{Q}{\mpl}\)^{-\beta}\comma
\end{equation}
where
\begin{eqnarray}
\beta &=& \frac{2\alpha}{2-\alpha}\comma\\
\psi &=& \frac{P_0}{\mpl^4}\[-1+\frac{2\chi}{\mpl^4}\]\comma\\
W(\psi) &=& \frac{(\psi\mpl^4+P_0)^2}{4P_0}\period
\end{eqnarray}
Provided that $X$ is constant during the scaling regime (so that $\psi$ is 
constant), the k-essence field behaves like an inverse power-law quintessence 
field in its well-known tracker regime. The transformation described in 
Section~\ref{dyn_equiv} allows us to find the appropriate power. We normalize 
the kinetic term and obtain
\begin{equation}
\Vti(\Qti)=\Vti_0\(\frac{\Qti}{\mpl}\)^{-\beta}\period
\end{equation}
where $\Vti_0=\psi^{\beta/2}(\beta/\alpha)^\beta(\psi\mpl^4+P_0)^2/4P_0$. 
Obviously, as soon as the scaling solution is modified (for instance at 
matter--radiation equality) $\psi$ evolves until the k-essence field enters a 
new scaling solution with a new quintessence-like behaviour. In \reffig{tracker} 
we show the evolution of $\psi$ for a cosmologically-realistic model, and also 
the evolution of the equation of state parameter $w$ for the k-essence field and 
for 
the almost equivalent quintessence model. The value of $\Vti_0$ has been chosen 
so that both models yield the same value of the equation of state parameter in 
the present epoch. Clearly, as long as $\psi$ does not remain exactly constant, 
the equivalence is not perfect and therefore the two models may in principle be 
distinguished from one another. However in practice we see in \reffig{tracker} 
that the evolution of $w$ is almost exactly the same in the two cases out to 
high redshift.

As a second example, consider the model described in Ref.~\cite{AMS1} given 
by
\begin{equation}
K(\phi)=\(\frac{\phi}{M_0}\)^{-2}\comma
\end{equation}
and
\begin{equation}
\begin{split}
\pti(X)=&\,M_0^4\Big[-2.01+2\sqrt{1+X/M_0^4}\\
        &+3\times10^{-17}\(X/M_0^4\)^3
         -10^{-24}\(X/M_0^4\)^4\Big]\comma
\end{split}
\end{equation}
where $M_0\equiv\sqrt{3/8\pi}\,\mpl$. The constants appearing in this expression 
are very specific, and it may be that there are simpler versions, but so far 
this is the best example that we know of in the literature that features an 
interesting property: the transition 
between an exact tracker regime and the domination of the k-essence field is 
triggered by matter--radiation equality and therefore, in a sense, the 
coincidence problem is solved. Again, for some ranges of $X$, contrary to what 
is assumed in the first part of Ref.~\cite{AMS1}, we can have $\varepsilon<0$, 
$\wk<-1$ or $\csk<0$ and $\wk$ and $\csk$ diverging at 
$X=X_w\approx2.1\times10^7\,M_0^4$ and $X=X_c\approx1.6\times10^7\,M_0^4$ 
respectively. Nevertheless, as long as we take $X\lesssim X_c$, we have 
$\varepsilon>0$, $\wk>-1$ and $\csk>0$ and, as explained in 
Section~\ref{k_essence}, we know that the field cannot cross this boundary.

Using the transformation described in Section~\ref{dyn_equiv} we find
\begin{equation}
V(Q)=e^{-2Q}\period
\end{equation}
Note that, since $\pti''(X)$ changes sign at $X\approx1.6\times10^6\,M_0^4$ and 
$X\approx1.5\times10^7\,M_0^4$, the Legendre transformation described in 
Section~\ref{dyn_equiv} cannot be computed. Nevertheless we still define $\psi$ 
as $\pti'(X)$, though $W$ can no longer be expressed as a function of $\psi$.

We have performed a numerical simulation in order to reproduce the result given 
in Ref.~\cite{AMS1}, that is to say a stable scaling solution during radiation 
domination which then evolves to k-essence domination after matter--radiation 
equality. As an aside, we find that the basin of attraction of this solution 
does not seem to be very large. Indeed, random initial conditions, even with a 
subdominant k-essence field, almost never lead to the desired solution, but 
instead to an early period of k-essence domination, or in some cases the 
solution even ceases to exist when, after a finite time, it reaches the 
singularity $X_\mathrm{c}$. However we will not pursue this further here, but 
leave it for a future investigation.

\begin{figure}[t]
\includegraphics[scale=0.38,angle=-90]{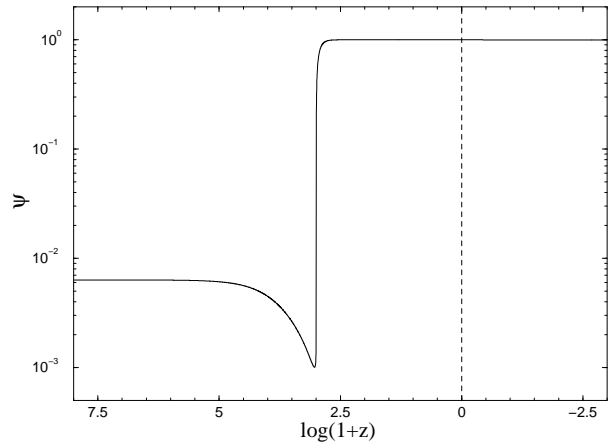}
\caption{Evolution of $\psi$ for the second example of a k-essence field 
described in Section~\ref{examples} in a cosmologically realistic case. The 
dramatic change in $\psi$ is triggered by the transition from radiation 
domination to matter domination.}
\label{psi}
\end{figure}

In \reffig{psi} we show the evolution of $\psi$ for this model. Clearly $\psi$ 
is almost constant during radiation domination, then evolves to another constant 
soon after matter--radiation equality, remaining nearly constant even during the 
transition to k-essence domination. Consequently, we know that during these two 
time 
periods the k-essence field behaves like a quintessence field. Renormalizing $Q$ 
in order to have a canonical kinetic term, we find two exponential potential 
quintessence models given by
\begin{equation}
\Vti(\Qti)\propto\exp(-\alpha_i\Qti)\qquad (i=1\ \mathrm{or}\ 2)\comma
\end{equation} 
with $\alpha_1\approx5.8\,\mpl^{-1}$ and $\alpha_2\approx73\,\mpl^{-1}$, which 
mimic the k-essence field during the two time periods described above 
respectively.

\begin{figure}[t]
\includegraphics[scale=0.38,angle=-90]{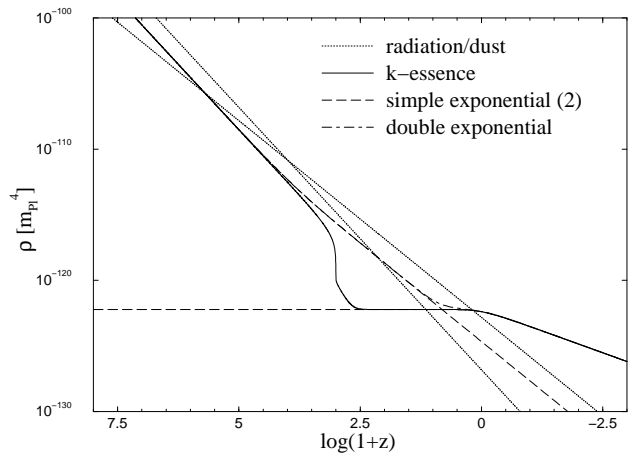}
\caption{Evolution of the energy density of radiation, dust-like matter and the 
second example of a k-essence model studied in Section~\ref{examples}. We also 
show the evolution of the energy density for two exponential quintessence models 
($\alpha_1\approx5.8\,\mpl^{-1}$ and $\alpha_2\approx72.8\,\mpl^{-1}$) and for 
the double exponential quintessence model obtained by adding both potentials.}
\label{rho}
\end{figure}

In order to mimic the k-essence field during these two time periods with only 
one quintessence model one may add both potentials to obtain a double 
exponential quintessence model~\cite{nunes}. In this case we have normalized the 
potential to fit the late-time behaviour of the k-essence field.
The evolution of the energy density for all these models is shown in 
\reffig{rho} and it is evident that the quintessence models closely mimic the 
k-essence field while $\psi$ is constant. Moreover, as expected, the double 
exponential model allows us to mimic the k-essence field for early times as well 
as for the late-time evolution of the universe.

\begin{figure}[t]
\includegraphics[scale=0.38,angle=-90]{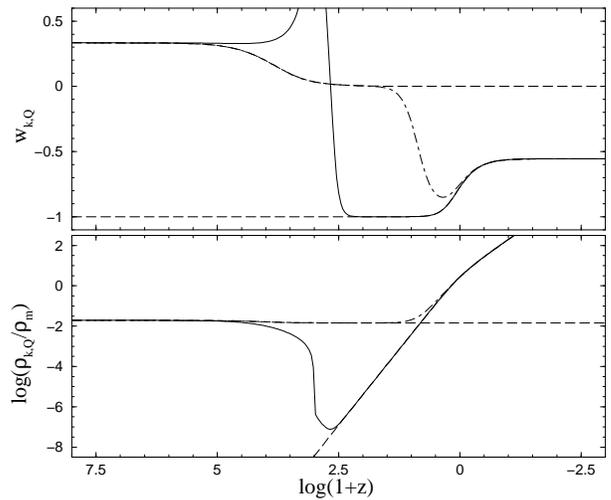}
\caption{Evolution of the equation of state $\wk$ (top) and the ratio 
$\rho_\mathrm{k}/\rho_\mathrm{m}$ (bottom) of the k-essence field as shown in 
Ref.~\cite{AMS1}. The quintessence fields plotted in \reffig{rho} are also 
shown, with the linestyles having the same meaning.}
\label{mukhanov}
\end{figure}

In \reffig{mukhanov} we reproduce the plots shown in Ref.~\cite{AMS1}, adding  
the quintessence models described here. Again, the quintessence models very 
closely mimic the k-essence model during the period for which they are 
equivalent; for instance the second exponential potential model 
gives indistinguishable evolution of $\rho_\mathrm{k}/\rho_\mathrm{m}$ and $w$ 
back to redshifts of several hundred (though note that this potential does not 
exhibit tracker behaviour).
Accordingly, observations studying the homogeneous evolution, such as supernovae 
observations, are not able to distinguish between such models.
For the double exponential model, the mimicking is not perfect during the recent 
past and therefore one might be able to tell the difference.

\section{Discussion}
\label{discussion}

Quintessence and k-essence are two attempts to explain, in terms of scalar 
fields, the current observation of an accelerating universe. Although they are 
similar in that they both involve the dynamics of light scalar fields, they 
differ in that quintessence relies on precise functional forms for the potential 
of the field, whereas k-essence derives its particular behaviour from the 
presence of non-canonical kinetic terms associated with the field. Given that 
both models attempt to explain the same observations, and that both models 
involve evolving scalar fields, a natural question that arises is whether it is 
possible to write one model in terms of the parameters of the other. 
In regimes where the effective equation of state parameter for the 
k-essence field becomes less than -1, such a relationship is not possible whilst 
maintaining conventional canonical kinetic terms for the quintessence fields, 
but for $\wk$ greater than -1 such equivalences may exist. 

In 
particular, in this paper we have addressed this question by attempting to 
rewrite k-essence models in terms of quintessence potentials, relating the two 
sets of fields through field redefinitions. Our results are intriguing; we have 
found a dynamically-equivalent action which has similarities with a canonical 
scalar field action and may be easier to study. We have examined two cases from 
the literature, and shown 
that during some regimes the homogeneous part of the k-essence field can behave 
exactly like a 
quintessence field, and have obtained exact equivalences in those cases. 

It could well prove impossible to differentiate between 
the two models, quintessence and k-essence, by measuring the evolution of the 
equation of state parameter. To distinguish between the models, it appears 
necessary to combine such studies with searches for subtle effects on the 
perturbations 
from the different sound speed in the two models \cite{sosp}.


\begin{acknowledgments}
M.M.~was supported by the Fondation Barbour, the Fondation Wilsdorf, the 
Janggen-P\"{o}hn-Stiftung and the ORS Awards Scheme, and A.R.L.~in part by the 
Leverhulme Trust. The work of MT is supported in part by the National Science 
Foundation under grant PHY-0094122. We thank Nicola Bartolo, Pier-Stefano 
Corasaniti and Paul Steinhardt for discussions.
\end{acknowledgments}


\appendix

\section{Extension to k-essence models with a general potential}
\label{extension}

In this section, we briefly extend our discussion to a more general 
model~\cite{MMOT} given by
\begin{equation}
S=\int\de^4x\sqrt{-g}\[\frac{\mpl^2}{16\pi}R
+K(\phi)\pti(X)-\bar{U}(\phi)\]\comma
\end{equation}
where we assume $K(\phi)>0$, $\pti(0)=0$ and 
$X=\frac12\nabla_\mu\phi\nabla^\mu\phi$. The pressure and the energy density are 
given by
\begin{eqnarray}
p &=& K(\phi)\pti(X)-\bar{U}(\phi)\comma\\
\varepsilon &=& K(\phi)\eti(X)+\bar{U}(\phi)\comma
\end{eqnarray}
where $\eti(X)$ is still defined by \refeq{eps_ti}. As usual, the equation of 
state parameter is given by $\wk=p/\varepsilon$ and the sound speed is still 
defined by \refeq{sound_speed}.

Now, we consider the action
\begin{equation}
\begin{split}
S=&\int d^4x\sqrt{-g}\bigg\{\frac{\mpl^2}{16\pi}R+K(\phi)\Big[\pti(X)\\
  &+\big(\frac12\nabla_{\mu}\phi\nabla^{\mu}\phi-\chi\big)\pti'(X)\Big]
  -\bar{U}(\phi)\bigg\}\period
  \end{split}
\end{equation}
and, using a method similar to that in Section~\ref{dyn_equiv}, and defining 
$U(Q)\equiv\bar{U}[\phi(Q)]$, it is possible to find an equivalent model 
described by the action
\begin{equation}
\begin{split}
S=&\int\de^4x\sqrt{-g}\bigg[-\frac{\mpl^2}{16\pi}R\\
&+\frac\psi2\nabla_\mu Q\nabla^\mu Q-V(Q)W(\psi)-U(Q)\bigg]\period
\end{split}
\end{equation}

The dynamical equations resulting from this action are
\begin{eqnarray}
\rho_\mathrm{Q\psi}&=&\frac{\psi}{2}\dot{Q}^2+V(Q)W(\psi) +U(Q)\comma\\
p_\mathrm{Q\psi}&=&\frac{\psi}{2}\dot{Q}^2-V(Q)W(\psi)-U(Q)\period
\end{eqnarray}
The Euler--Lagrange equation for $Q$ is given by
\begin{equation}
\psi\ddot{Q}+(3H\psi+\dot{\psi})\dot{Q}+V'(Q)W(\psi)+U'(Q)=0\comma
\end{equation}
and the constraint equation for $\psi$ is still given by Eq.~(\ref{constraint}).

Canonically normalizing $Q$ via $\Qti\equiv\sqrt{\abs{\psi}}Q$, the equation of 
motion for the field $\Qti$ is then
\begin{equation}
\begin{split}
\ddot{\Qti}&+3H\dot{\Qti}\pm\Vti'(\Qti)W(\psi)\pm\Uti'(\Qti)\\
           &-\frac{3H\Qti\dot{\psi}}{2\psi}
            +\frac{\Qti\dot{\psi}^2}{4\psi^2}
            -\frac{\Qti\ddot{\psi}}{2\psi}=0\comma
\end{split}
\end{equation}
where ``$\pm$'' stands for the sign of $\psi$ and 
\begin{eqnarray}
\Vti(\Qti) &\equiv& V\big(\Qti/\sqrt{\abs{\psi}}\big)\comma\\
\Uti(\Qti) &\equiv& U\big(\Qti/\sqrt{\abs{\psi}}\big)\period
\end{eqnarray}
In addition, the constraint equation transforms to
\begin{equation}
\dot{\Qti}^2-2\abs{\psi}W'(\psi)\Vti(\Qti)-\frac{\Qti\dot{\Qti}\dot{\psi}}{\psi}
+\frac{\Qti^2\dot{\psi}^2}{4\psi^2}=0\period
\end{equation}

In the limit in which the approximation $\psi\approx\mathrm{constant}$ applies, 
the equation of motion becomes
\begin{equation}
\ddot{\Qti}+3H\dot{\Qti}\pm\Vti'(\Qti)W(\psi)\pm\Uti'(\Qti)=0\comma
\end{equation} 
and the constraint equation simplifies significantly to become
\begin{equation}
\dot{\Qti}^2-2\abs{\psi}W'(\psi)\Vti(\Qti)=0\period
\end{equation} 
Thus, even in this extended class of models, if $\psi$ remains constant during 
some time period, the dynamics will be equivalent to that of a quintessence 
model or a phantom model for this period.

As a simple example, consider the k-essence example of Ref.~\cite{MMOT}, in 
which the 
equation state $w<-1$ is obtained. This has $p(X)=P_0\[\exp(-\alpha 
X/\mpl^4)-1\]$, where $\alpha>0$ and $P_0>0$, and unspecified functions for 
$K(\phi)$ and ${\bar U}(\phi)$. This means that the translation to our new 
action is effected by
\begin{eqnarray}
\psi &=& -\frac{\alpha P_0}{\mpl^4}\exp\(-\alpha\chi/\mpl^4\)\comma\\
W(\psi) &=& P_0+\frac{\psi\mpl^4}{\alpha}
\[1-\ln\(-\frac{\psi\mpl^4}{\alpha P_0}\)\]\period
\end{eqnarray}
Note that $-\alpha P_0/\mpl^4<\psi<0$ so that this model does indeed correspond 
to $w<-1$. The condition that the energy density be positive becomes
\begin{equation}
\frac{U(Q)}{V(Q)}>
\frac{\psi\mpl^4}{\alpha}\[2\ln\(-\frac{\psi\mpl^4}{\alpha P_0}\)-1\]-P_0\period
\end{equation}
Further, a necessary condition for stability of the theory is that the sound 
speed be positive $c_s^2>0$. This condition becomes
\begin{equation}
\psi<-\frac{\alpha P_0}{\mpl^4}\exp\(-1/2\)\comma
\end{equation}
which is satisfied for $X<X_\mathrm{c}=\mpl^4/2\alpha$. Therefore we assume 
the theory to be valid until some cut-off below~$X_\mathrm{c}$.



\begin{thebibliography}{}
\bibitem{acce} A. G. Riess et al., Astron. J. \textbf{116}, 1009 (1998),
        \texttt{astro-ph/9805201}; P. Garnavich et al., Astrophys. J.
        \textbf{509}, 74
        (1998), \texttt{astro-ph/9806396}; S. Perlmutter et al., Astrophys. J.
        \textbf{517}, 565 (1998), \texttt{astro-ph/9812133}; G. Efstathiou 
        et al., Mon. Not. Roy. Ast. Soc. \textbf{330}, L29 (2002), 
        \texttt{astro-ph/0109152}.
\bibitem{quin} B. Ratra and P. J. E. Peebles, Phys. Rev. D\textbf{37}, 3406 
        (1988); E. J. Copeland, A. R. Liddle, and D. Wands, Ann. N. Y. Acad. 
	Sci. \textbf{688}, 647 (1993); R. R. Caldwell, R. Dave, and P. J. 
	Steinhardt, Phys. Rev. Lett. \textbf{80}, 1582 (1998),
	\texttt{astro-ph/9708069}; P. G. Ferreira and M. Joyce, Phys. Rev. 
	D\textbf{58}, 023503 (1998), \texttt{astro-ph/9711102}; E. J. 
	Copeland, A. R. Liddle, and D. Wands, Phys. Rev. 
	D\textbf{57}, 4686 (1998), \texttt{gr-qc/9711068}; I. Zlatev, L. Wang,
	and P. J. Steinhardt, Phys. Rev. Lett. \textbf{82}, 896 (1999),
        \texttt{astro-ph/9807002}; A. R. Liddle and R. J. Scherrer, Phys. 
        Rev. D\textbf{59}, 023509 (1999), \texttt{astro-ph/9809272}; V. Sahni
        and A. Starobinsky, Int. J. Mod. Phys. D\textbf{9}, 373 (2000),
        \texttt{astro-ph/9904398}.
\bibitem{trac} C. Wetterich, Nucl. Phys. \textbf{B302}, 668 (1988); I. Zlatev,
        L. Wang, and P. J. Steinhardt, Phys. Rev. Lett. \textbf{82}, 896 (1999),
        \texttt{astro-ph/9807002}.
\bibitem{ADM} C. Armend\'ariz-Pic\'on, T. Damour, and V. Mukhanov, Phys. Lett. 
        B\textbf{458}, 219 (1999), \texttt{hep-th/9904075}.
\bibitem{GM} J. Garriga and V. Mukhanov, Phys. Lett. B\textbf{458}, 219 (1999),
        \texttt{hep-th/9904176}.
\bibitem{COY} T. Chiba, T. Okabe, and M. Yamaguchi, Phys. Rev. D\textbf{62}, 
        023511 (2000), \texttt{astro-ph/9912463}.
\bibitem{AMS1} C. Armend\'ariz-Pic\'on, V. Mukhanov, and P. J. Steinhardt, Phys. 
        Rev. Lett. \textbf{85}, 21 (2000), \texttt{astro-ph/0004134}.
\bibitem{AMS2} C. Armend\'ariz-Pic\'on, V. Mukhanov, and P. J. Steinhardt, Phys. 
        Rev. D\textbf{63}, 103510 (2001), \texttt{astro-ph/0006373}.
\bibitem{tach} A. Sen, JHEP 0204, 048 (2002), \texttt{hep-th/0203211};
        A. Sen, Mod. Phys. Lett. A\textbf{17}, 1797 (2002), 
        \texttt{hep-th/0204143}; G. W. Gibbons, Phys. Lett. B\textbf{537}, 1
        (2002), \texttt{hep-th/0204008}; J. S. Bagla, H. K. Jassal, and T.
        Padmanabhan, \texttt{hep-th/0212198}. 
\bibitem{MMOT} A. Melchiorri, L. Mersini, C. J. \"Odman, and M. Trodden,
        \texttt{astro-ph/0211522}.
\bibitem{CHT} S. M. Carroll, M. Hoffman, and M. Trodden,
        \texttt{astro-ph/0301273}.
\bibitem{sosp} J. Erickson, R. R. Caldwell, P. J. Steinhardt, V. Mukhanov, and
        C. Armend\'ariz-Pic\'on, Phys. Rev. Lett. \textbf{88}, 121301 (2001),
        \texttt{astro-ph/0112438}; S. DeDeo, R. R. Caldwell, and P. J.
        Steinhardt, \texttt{astro-ph/0301284}.
\bibitem{KS} H. Kodama and M. Sasaki, Prog. Theor. Phys. Suppl. \textbf{78},
        1 (1984).
\bibitem{phan} R. R. Caldwell, Phys. Lett. B\textbf{545}, 23 (2002),
        \texttt{astro-ph/9908168}.
\bibitem{nunes} T. Barreiro, E. Copeland, and N. Nunes, 
	Phys. Rev. \textbf{D}61, 127301 (2000), \texttt{astro-ph/9910214}. 
\end{thebibliography}
\end{document}